\documentclass[aps,prb,twocolumn,showpacs,amsmath,amssymb,floatfix,superscriptaddress]{revtex4-1}
\usepackage{graphicx}
\usepackage[bookmarks,bookmarksopen,bookmarksnumbered,colorlinks,linkcolor=red,linktocpage,citecolor=blue,urlcolor=cyan,pdfpagemode=UseOutline]{hyperref}

\def\ben{\begin{equation}}
\def\een{\end{equation}}
\def\benu{\begin{enumerate}}
\def\enu{\end{enumerate}}
\def\sss{\scriptscriptstyle\rm}
\def\s{_{\sss S}}
\def\xc{_{\sss XC}}
\def\br{{\bf r}}

\newcommand{\intd}{\mathrm{d}}
\providecommand{\abs}[1]{\left|#1\right|}
\newcommand{\parref}[1]{(\ref{#1})}
\newcommand{\vect}[1]{\mathbf{#1}}
\newcommand{\matelem}[3]{\left\langle #1 \left| #2 \right| #3 \right\rangle}
\providecommand{\bra}[1]{\left< #1 \right|}
\providecommand{\ket}[1]{\left| #1 \right>}
\DeclareMathOperator{\tr}{tr}

\def\bk{\mathbf{k}}
\def\xcsigma{_{{\sss XC},\sigma}}
\def\ssigma{_{{\sss S},\sigma}}
\def\taudep{^{\tau-\text{dep}}}
\def\tauELP{^{\tau\text{ELP}}}
\def\NL{^\text{gKS}}

\def\BZ{_\text{BZ}}
\def\OmegaBZ{\Omega_{\sss BZ}}

\begin{document}

\title{More-Realistic Band Gaps from Meta-Generalized Gradient Approximations: Only in a Generalized Kohn-Sham Scheme}
\author{Zeng-hui Yang}
\affiliation{Department of Physics, Temple University, Philadelphia, PA 19122, USA}
\author{Haowei Peng}
\affiliation{Department of Physics, Temple University, Philadelphia, PA 19122, USA}
\author{Jianwei Sun}
\affiliation{Department of Physics, Temple University, Philadelphia, PA 19122, USA}
\author{John P. Perdew}
\affiliation{Department of Physics, Temple University, Philadelphia, PA 19122, USA}
\date{\today}

\begin{abstract}
Unlike the local density approximation (LDA) and the generalized gradient approximation (GGA), calculations with meta-generalized gradient approximations (meta-GGA) are usually done according to the generalized Kohn-Sham (gKS) formalism. The exchange-correlation potential of the gKS equation is non-multiplicative, which prevents systematic comparison of meta-GGA bandstructures to those of the LDA and the GGA. We implement the optimized effective potential (OEP) of the meta-GGA for periodic systems, which allows us to carry out meta-GGA calculations in the same KS manner as for the LDA and the GGA. We apply the OEP to several meta-GGAs, including the new SCAN functional [Phys. Rev. Lett. 115, 036402 (2015)]. We find that the KS gaps and KS band structures of meta-GGAs are close to those of GGAs. They are smaller than the more realistic gKS gaps of meta-GGAs, but probably close to the less-realistic gaps in the band structure of the exact KS potential, as can be seen by comparing with the gaps of the EXX+RPA OEP potential. The well-known grid sensitivity of meta-GGAs is much more severe in OEP calculations.
\end{abstract}

\maketitle

\section{Introduction}
Semiconductor devices play an important role in modern technologies, and the rapid development of electronic structure theory methods has made computational design of such devices possible. The band gap and the band structure are undoubtly the most important properties of semiconductors, since these are the properties that distinguish semiconductors from other periodic systems\cite{YC10}. Computational evaluation of the band gap and the band structure is thus a topic of active research.

The fundamental band gap is a ground-state property, and it is defined as $E_g=I-A$, where I is the ionization energy, and A is the electron affinity. I and A are ground-state energy differences. $E_g$ is also an excited-state property since it is the unbound limit of the exciton series. $E_g$ is very difficult to calculate for periodic systems, since there is no systematic way of adding/removing one electron to/from the solid in a periodic calculation, and the bulk limit can only be approached by the calculation of very big clusters. Many-body methods such as the GW method\cite{H65} calculate $E_g$ and the quasiparticle band structure accurately, but the computational cost is high.

The density-functional theory (DFT)\cite{HK64,KS65,FNM03} is a formally exact electronic structure method for the ground-state energy and electron density with an excellent balance of accuracy and computational efficiency, which is achieved by mapping the real interacting system to a fictitious Kohn-Sham (KS) system of non-interacting electrons with a multiplicative effective exchange-correlation (xc) potential (the functional derivative of the exchange-correlation energy with respect to the density). The exact Kohn-Sham potential yields the exact density but \emph{not} the exact quasi-particle band structure and gap. Though the exact energy functional of the DFT is unknown, there exists a plethora of approximations, which has been ordered into the `Jacob's ladder'\cite{PS01} hierarchy. The first and the second rungs of the Jacob's ladder are the local density approximation (LDA) and the generalized gradient approximation (GGA), and they severely underestimate the fundamental gap in periodic systems. For periodic systems, KS DFT cannot calculate $E_g$ from its definition, and one commonly approximates $E_g$ with the KS gap $E_g^\text{KS}=\epsilon^\text{KS}_\text{LUMO}-\epsilon^\text{KS}_\text{HOMO}$, where $\epsilon^\text{KS}_\text{HOMO}$ and $\epsilon^\text{KS}_\text{LUMO}$ are the KS orbital energies of the highest occupied orbital and of the lowest unoccupied orbital, respectively. However, $E_g^\text{KS}$ is not equal to $E_g$ even with the exact functional, due to the derivative discontinuity (DD)\cite{PPLB82}. The band gap problem has been an obstacle in the application of DFT to periodic systems.

The generalized Kohn-Sham (gKS)\cite{SGVM96} scheme is a different formulation of the DFT, which allows a non-multiplicative but still Hermitian xc potential operator. The gKS gap $E_g^\text{gKS}=\epsilon^\text{gKS}_\text{LUMO}-\epsilon^\text{gKS}_\text{HOMO}$ can be a better approximation to $E_g$ than is the KS gap\cite{C99}. The third rung of the Jacob's ladder, the meta-generalized gradient approximation (meta-GGA), is commonly implemented in the gKS scheme according to the method of Neumann, Nobes and Handy (also denoted as gKS in this paper)\cite{NNH96}. The gKS meta-GGA band gap of periodic systems improves\cite{ZT09} over the KS GGA gaps as expected. In this work we find that, with the recently proposed strongly constrained and appropriately normed (SCAN) functional\cite{SRP15,SRZS16}, the gKS meta-GGA gap corrects about 20\%$\sim$50\% of the difference between the experimental fundamental gap and the GGA KS gap.

Due to the restriction in the functional form of GGA, a GGA cannot perform well for finite systems and periodic systems at the same time.\cite{PRCV08} On the other hand, the functional form of meta-GGA can satisfy more exact conditions and has a wider range of applicability than the GGA form. The SCAN functional is a non-empirical functional that satisfies all known exact constraints appropriate to a semilocal functional, and is expected to perform well for systems with very different kinds of bonds. The computational accuracies of the SCAN functional for many properties improve over those of the GGAs, with only marginal increase of computational cost.\cite{SRP15} We find that SCAN also improves band gaps, but the comparsion is between meta-GGA gKS gaps and GGA KS gaps, which are not the same quantity. It is unclear whether the KS gap itself is improved, or just the gKS gap is improved. One needs to do meta-GGA calculations in the KS scheme to be able to have a systematic comparsion of the KS band gaps between meta-GGA and GGA.

In the gKS formalism for an orbital functional such as a meta-GGA or hybrid functional, we find the ``optimal variational potential'', a non-multiplicative self-adjoint operator that minimizes the energy functional with respect to the orbitals. This potential is a differential operator for a meta-GGA and an integral operator for a hybrid functional. In the KS formalism for an orbital functional, we find the optimized effective potential (OEP)\cite{SH53,TS76}, the multiplicative xc potential that minimizes the energy. The OEP meta-GGA yields the KS gap of the meta-GGA, which can be compared directly with the GGA KS gap. With the OEP, the meta-GGA xc potential also becomes comparable with that of the GGA. The OEP meta-GGA has been studied only in finite systems previously.\cite{AK03,EH14}

In this work, we provide the first study of the OEP meta-GGA in periodic systems. We find that the meta-GGA KS gap is not significantly improved over the GGA KS gaps, and the band structure of the OEP meta-GGA is very close to that of the GGA (and presumably to that of the exact KS potential). The xc potential of the meta-GGA has more details than that of the GGA, but the change is not as big as the change between the LDA and the GGA. Though the gKS meta-GGA is known to be sensitive to the real-space grid used in the calculation, we find that the OEP meta-GGA has worse grid sensitivity. The reason for the grid sensitivity is discussed in this work.

Within the gKS implementation of the meta-GGA form, it is possible to fit to energy gaps of solids\cite{TB09}, and the result can be useful for the prediction of gaps\cite{KG15}. But the fact that a property can be fitted in DFT is not evidence that it should be, or that other and more appropriate properties will not deteriorate as a result.

\section{OEP meta-GGA in periodic systems}
\subsection{Theory}
\label{sect:theory}
The meta-GGA xc energy functional has the form
\ben
\begin{split}
&E\xc[n_\uparrow,n_\downarrow]\\
&=\int\intd^3r e\xc[n_\uparrow(\br),n_\downarrow(\br),\nabla n_\uparrow(\br),\nabla n_\downarrow(\br),\tau_\uparrow(\br),\tau_\downarrow(\br)],
\end{split}
\een
where $e\xc$ is the xc energy density, and $n_\sigma$ and $\tau_\sigma$ are the spin density and the kinetic energy density of spin $\sigma$ respectively. $n_\sigma$ and $\tau_\sigma$ are
\begin{align}
n_\sigma(\br)&=\frac{1}{\OmegaBZ}\sum_i\int\limits\BZ\intd^3k\;\theta_{i\sigma}(\bk) \abs{\psi_{i\bk\sigma}(\br)}^2,\\
\tau_\sigma(\br)&=\frac{1}{2\OmegaBZ}\sum_i\int\limits\BZ\intd^3k\; \theta_{i\sigma}(\bk)\abs{\nabla\psi_{i\bk\sigma}(\br)}^2,
\end{align}
where $\OmegaBZ$ is the volume of the first Brillouin zone (BZ), $i$ is the band index, $\psi_{i\bk\sigma}$ is the KS orbital normalized in one unit cell, and $\theta_{i\sigma}(\bk)=\theta[E_\text{F}-\epsilon_{i\bk\sigma}]$ is the Heaviside step function, with $E_\text{F}$ being the Fermi energy, and $\epsilon_{i\bk\sigma}$ being the KS orbital energy. The non-interacting kinetic energy density of the Kohn-Sham orbitals is, like the orbitals themselves, a functional of the electron density. The OEP of the meta-GGA is defined by the functional derivative of the xc energy with respect to the density, which is
\ben
\begin{split}
&v\xcsigma(\br)=\frac{\delta E\xc}{\delta n_\sigma(\br)}\\
&=\frac{\partial e\xc}{\partial n_\sigma}(\br)-\nabla\cdot\left[\frac{\partial e\xc}{\partial\nabla n_\sigma}(\br)\right]+\int\limits\intd^3r'\;\frac{\partial e\xc}{\partial\tau_\sigma}(\br')\frac{\delta\tau_\sigma(\br')}{\delta n_\sigma(\br)}.
\label{eqn:vxcdef}
\end{split}
\een
Eq. \parref{eqn:vxcdef} can be partitioned into a GGA-like part and a $\tau$-dependent part:
\ben
v\xcsigma(\br)=v\xcsigma^\text{GGA}(\br)+v\xcsigma\taudep(\br).
\een
$v\xcsigma^\text{GGA}$ is a multiplicative potential, but $v\xcsigma\taudep$ does not have a closed form. $v\xcsigma\taudep$ can be written as
\ben
\begin{split}
v\xcsigma\taudep(\br)=&\frac{1}{\OmegaBZ}\sum_{i}\int\limits\BZ\intd^3k\iint\limits_{V_\text{cell}}\intd^3r'\intd^3r''\;\frac{\partial e\xc}{\partial\tau_\sigma}(\br')\\
&\times\left[\frac{\delta\tau_\sigma(\br')}{\delta\psi_{i\bk\sigma}(\br'')}\frac{\delta\psi_{i\bk\sigma}(\br'')}{\delta n_\sigma(\br)}+\text{c.c.}\right].
\end{split}
\label{eqn:v2}
\een
On one hand, the gKS meta-GGA potential operator is obtained by multiplying both side of Eq. \parref{eqn:v2} by $\delta n_\sigma(\br)/\delta\psi_{i\bk'\sigma'}(\br_1)+\text{c.c}$ and integrating over $\br$:
\ben
\begin{split}
v\xcsigma\taudep(\br)\psi_{i\bk\sigma}(\br)&=-\frac{1}{2}\nabla\cdot\left[\frac{\partial e\xc}{\partial\tau_\sigma}(\br)\nabla\psi_{i\bk\sigma}(\br)\right]\\
&\equiv \hat{v}\xcsigma\NL\psi_{i\bk\sigma}(\br).
\end{split}
\label{eqn:nonlocal}
\een
On the other hand, inserting
\ben
\frac{\delta\psi_{i\bk\sigma}(\br'')}{\delta n_\sigma(\br)}=\int\limits_{V_\text{cell}}\intd^3r_1\;\frac{\delta\psi_{i\bk\sigma}(\br'')}{\delta v\ssigma(\br_1)}\frac{\delta v\ssigma(\br_1)}{\delta n_\sigma(\br)},
\een
where $v\s$ is the KS potential, into Eq. \parref{eqn:v2} yields the OEP integral equation for $v\xcsigma\taudep$:
\ben
\begin{split}
&\int\limits_{V_\text{cell}}\intd^3r\; v\xcsigma\taudep(\br)\frac{\delta n_\sigma(\br)}{\delta v\ssigma(\br_1)}+\text{c.c.}\\
=&\frac{1}{\OmegaBZ}\sum_{i}\int\limits\BZ\intd^3k\iint\limits_{V_\text{cell}}\intd^3r'\intd^3r''\frac{\partial e\xc}{\partial\tau_\sigma}(\br')\frac{\delta\tau_\sigma(\br')}{\delta\psi_{i\bk\sigma}(\br'')}\frac{\delta\psi_{i\bk\sigma}(\br'')}{\delta v\ssigma(\br_1)}+\text{c.c.}.
\end{split}
\label{eqn:OEPint}
\een
Though a direct solution for the OEP is possible\cite{IHB99,EH14}, it is both computationally expensive and numerically unstable. The common practice is to approximate the solution of Eq. \parref{eqn:OEPint}. Ref. \onlinecite{AK03} employed the local-Hartree-Fock (LHF)\cite{DG01} approximation to derive approximated $v\xcsigma\taudep$ of meta-GGA for finite systems. Here we do the derivation with the effective local potential (ELP)\cite{SSD06,ISSD07} approximation. The ELP is equivalent to the LHF and the common-energy-denominator approximation (CEDA)\cite{GB01,GGB02}, and the Krieger-Li-Iafrate (KLI)\cite{KLI92} approximation is their simplification.

The ELP minimizes the matrix norm of the commutator $S_\sigma=[\hat{D}_\sigma,\hat{\rho}_\sigma]$, where $\hat{D}_\sigma=v\xcsigma\taudep-\hat{v}\xcsigma\NL$, and $\hat{\rho}_\sigma=\frac{1}{\OmegaBZ}\sum_i\int\intd^3k\;\theta_{i\sigma}(\bk)\ket{\psi_{i\bk\sigma}}\bra{\psi_{i\bk\sigma}}$ is the spin density matrix. $\hat{\rho}_\sigma$ becomes valid for both the KS and the gKS system when $S_\sigma=0$. $S_\sigma$ is minimized when $\delta S_\sigma/\delta v\xcsigma\taudep(\br)=0$, which yields the ELP approximation of $v\xcsigma\taudep$:
\ben
\begin{split}
&v\xcsigma\tauELP(\br)=\\
&\frac{1}{2\OmegaBZ n_\sigma(\br)}\int\limits\BZ\intd^3k\Big\{\sum_i\theta_{i\sigma}(\bk)\psi^*_{i\bk\sigma}(\br)\hat{v}\xcsigma\NL\psi_{i\bk\sigma}(\br)\\
&+\frac{1}{\OmegaBZ}\int\limits\BZ\intd^3k'\sum_{ij}\theta_{i\sigma}(\bk)\theta_{j\sigma}(\bk')\psi^*_{i\bk\sigma}(\br)\psi_{j\bk'\sigma}(\br)\\
&\times\matelem{\psi_{j\bk'\sigma}}{\hat{D}_\sigma}{\psi_{i\bk\sigma}}\Big\}+\text{c.c.}
\end{split}
\label{eqn:elppot}
\een
The ELP approximates the OEP in a least-squares sense\cite{ISSD07}.

Ref. \onlinecite{ISSD07} shows that the ELP approximation is equivalent to the CEDA, which reduces to the KLI approximation by setting $i=j$ in Eq. \parref{eqn:elppot}\cite{GB01}. If only the first term of Eq. \parref{eqn:elppot} is kept, one obtains the so-called Slater approximation\cite{S51}.

\subsection{Implementation in the BAND code}
The most efficient methods for calculating periodic systems are pseudopotential methods with planewave basis sets, with the widely used PAW method as an extension\cite{B94, KJ99}, but the OEP is hard to implement in such codes. The sum over bands in Eq. \parref{eqn:elppot} also includes the core bands, which are not directly available in pseudopotential codes. Furthermore, the OEP is inherently incompatible with the PAW formalism. The charge density in the PAW formalism is composed of the pseudo-charge-density, the on-site all-electron density, and the on-site pseudo-charge-density\cite{B94, KJ99}, and the xc energy and potential are partitioned into three parts accordingly, with each part only depending on the corresponding part of the density. However, the OEP of meta-GGA is hard to partition this way. Due to these obstacles, we implement the OEP meta-GGA in the BAND code\cite{VB91, WB91, FPV13, FPLV14, BAND2014}.

BAND is an all-electron DFT code for periodic systems. It uses atom-centered numerical functions as the basis set. All the real-space integrations are done numerically on a Becke fuzzy-cell grid\cite{B88}. The BZ is discretized with the commonly used Monkhorst-Pack grid\cite{MP76}, so the $\bk$-integrations of Eq. \parref{eqn:elppot} become sums over $\bk$-points in the irreducible Brillouin zone (IBZ):
\ben
\frac{1}{\OmegaBZ}\int\intd^3k\;g_{i\sigma}(\bk)\theta_{i\sigma}(\bk)=\frac{1}{N_\text{G}}\sum_{\tilde{\bk}}^\text{IBZ}\sum_{\hat{S}} w_{\tilde{\bk}}f_{i\tilde{\bk}\sigma}g_{i\sigma}(\hat{R}^{-1}\tilde{\bk})
\label{eqn:occ}
\een
where $g_{i\sigma}(\bk)$ is the contribution to the integral from the orbital $\psi_{i\bk\sigma}$, $f_{i\tilde{\bk}\sigma}$ is the occupation number, $w_{\tilde{\bk}}$ is the weight of the $\bk$-point times the $\bk$-space integration weight in BAND\cite{WB91}, $N_\text{G}$ is the total number of symmetry operations, $\hat{S}=\{\hat{R},\vect{t}\}$ is the space-group symmetry operator, and $\hat{R}$ is the rotational part of $\hat{S}$.

The $\hat{D}_\sigma$ matrix elements in Eq. \parref{eqn:elppot} vanish for $\bk\ne\bk'$, since $\hat{D}_\sigma$ has the perodicity of the unit cell. Eq. \parref{eqn:elppot} is implemented in the BAND code as
\ben
\begin{split}
&v\xcsigma\tauELP(\br)=\\
&\frac{1}{N_\text{G}n_\sigma(\br)}\sum_{\tilde{\bk}}^\text{IBZ}\sum_{\hat{S}}\sum_i w_{\tilde{\bk}} f_{i\tilde{\bk}\sigma}\\
&\times\Re\Bigg\{-\frac{1}{2}\psi_{i\tilde{\bk}\sigma}^*(\br)\left[\nabla_\br\frac{\partial e\xc}{\partial\tau_\sigma}(\br)\cdot\hat{R}^{-1}\nabla_{\br'}\psi_{i\tilde{\bk}\sigma}(\br')|_{\br'=\hat{S}\br}\right.\\
&\left.+\frac{\partial e\xc}{\partial\tau_\sigma}(\br)\tr\left.\left\{\hat{R}^{-1}\cdot\left[\nabla_{\br'}\otimes\nabla_{\br'}\psi_{i\tilde{\bk}\sigma}(\br')\right]\cdot\hat{R}\right\}\right|_{\br'=\hat{S}\br}\right]\\
&+\sum_j f_{j\tilde{\bk}\sigma}\psi_{i\tilde{\bk}\sigma}^*(\hat{S}\br)\psi_{j\tilde{\bk}\sigma}(\hat{S}\br)(I_{ij\tilde{\bk}\sigma}-J_{ij\tilde{\bk\sigma}})\Bigg\},
\end{split}
\label{eqn:working}
\een
where $I$ and $J$ are
\begin{align}
I_{ij\tilde{\bk}\sigma}&=\int\limits_{V_\text{cell}}\intd^3r\;\psi_{i\tilde{\bk}\sigma}^*(\br)v\xcsigma\tauELP(\br)\psi_{j\tilde{\bk}\sigma}(\br),\\
\label{eqn:Iint}
J_{ij\tilde{\bk}\sigma}&=\frac{1}{2}\int\limits_{V_\text{cell}}\intd^3r\; \frac{\partial e\xc}{\partial\tau_\sigma}(\br)\nabla\psi_{i\tilde{\bk}\sigma}^*(\br)\cdot\nabla\psi_{j\tilde{\bk}\sigma}(\br),
\end{align}
and $\nabla_{\br'}\otimes\nabla_{\br'}\psi_{i\tilde{\bk}\sigma}(\br')$ denotes the Hessian matrix of $\psi_{i\tilde{\bk}\sigma}$. The integrals are done as discrete sums on the real-space grid. The $I_{ij\tilde{\bk}\sigma}$ depend on $v\xcsigma\tauELP$, which is unknown beforehand. They can be determined by solving linear equations\cite{AK03}, but this is costly for perodic systems. Instead, we determine the $I$ integrals and $v\xcsigma\tauELP$ iteratively by the following steps:
\benu
\item Set the $I$ integrals to 0.
\item Calculate $v\xcsigma\tauELP$ with Eq. \parref{eqn:working} using the $I$ integrals of the last iteration.
\item Calculate the $I$ integrals using the new $v\xcsigma\tauELP$.
\item Repeat step 2 and 3 if $\int\intd^3r \abs{v\xcsigma^{\tau\text{ELP}(i)}(\br)-v\xcsigma^{\tau\text{ELP}(i-1)}(\br)}>\mathcal{T}$, where $i$ is the iteration count, and $\mathcal{T}$ is the error tolerance.
\enu
$v\xcsigma\tauELP$ is only determined up to an additive constant $c$, since $v\xcsigma\tauELP+c$ yields the same band structure. $c$ has to be fixed for determining the convergence. The $I$ integrals contain $v\xcsigma\tauELP$, so their values include $c$. We fix $c$ by requiring that the term of $v\xcsigma\tauELP$ containing the $I$ integrals averages to 0, i.e. the constant $c$ is determined by
\begin{multline}
\frac{1}{V_\text{cell}}\int\limits_{V_\text{cell}}\intd^3r\frac{1}{N_Gn_\sigma(\br)}\sum_{\tilde{\bk}}^\text{IBZ}\sum_{\hat{S}}\sum_i w_{\tilde{\bk}} f_{i\tilde{\bk}\sigma}\sum_j f_{j\tilde{\bk}\sigma}\\
\times\psi_{i\tilde{\bk}\sigma}^*(\hat{S}\br)\psi_{j\tilde{\bk}\sigma}(\hat{S}\br)\left(I_{ij\tilde{\bk}\sigma}+c\delta_{ij}\right)=0
\end{multline}
In practice, the iterative procedure converges after 200 iterations on average for $\mathcal{T}=10^{-10}$.

Ref. \onlinecite{GGB02} finds that the total energies and orbital energies obtained from CEDA and KLI are very close for atoms and molecules. We find that the same is true for periodic systems. The ELP is equivalent to the CEDA\cite{ISSD07}, and its total energy is always lower than that of the KLI, but the total energy differences and the KS gap differences of the materials in Table \ref{table:gap} between that of the ELP and that of the KLI are at most 1 meV. Since the improvement of the ELP over the KLI is insignificant, we only report the KLI results in the following.

\section{Results}
We calculate the KS and gKS gaps of 20 semiconductors and insulators with the SCAN, meta-GGA made simple 2 (MS2)\cite{SXR12,SHXB13}, meta-GGA made very simple (MVS)\cite{SPR14} and TPSS\cite{TPSS03} meta-GGAs and the PBE\cite{PBE96} GGA, and the results are collected in Table \ref{table:gap}. Ge, CdO, and InN are not listed since their calculated gaps vanish. All the calculations are done with the TZ2P basis set\cite{LB03}. A $9\times9\times9$ Monkhorst-Pack $\bk$-grid\cite{MP76} is used for most of the materials except InN, CdS and CdSe, for which a $9\times9\times5$ $\bk$-grid is used. We find that the OEP meta-GGA is sensitive to the real-space grid, and large grids are used to ensure convergence. The details of the grid problem are discussed in Section \ref{sect:grid}. Though GGA and gKS meta-GGA converge properly with smaller grids, we use the same grid as the OEP meta-GGA, so that the results are comparable.

The scalar relativistic effect is included in the calculations by the ZORA\cite{PLSB97} method. We ignore the spin-orbit coupling since the effect is small (0.03 eV for InP). The relativistic effect is implicitly included through the pseudopotential in planewave codes, but it needs to be explicitly included in BAND. We find that the scalar relativistic effect has a big impact on the KS and gKS gaps, since it shifts down the orbital energies of $s$-type bands, such as the lowest conduction band of GaAs\cite{BC85}. The change in the KS and gKS gaps due to the scalar relativistic effect can be as big as 0.7 eV (GaAs).

\begin{figure}[htbp]
\includegraphics[width=0.95\columnwidth]{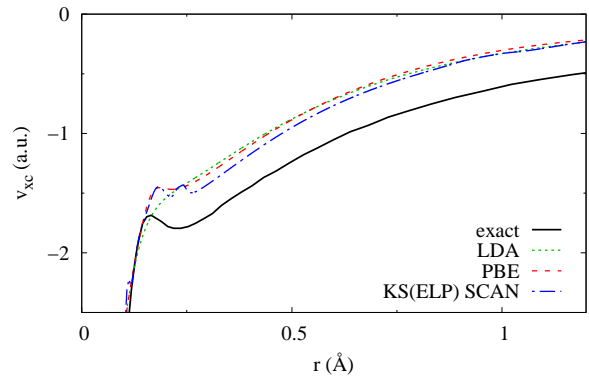}
\caption{(color online) Comparison of $v\xc$'s of the Ne atom for SCAN, LDA, and GGA. The exact $v\xc$ is from Ref. \onlinecite{RS12}.}
\label{fig:NeVxc}
\end{figure}

\begin{figure}[htbp]
\includegraphics[width=0.95\columnwidth]{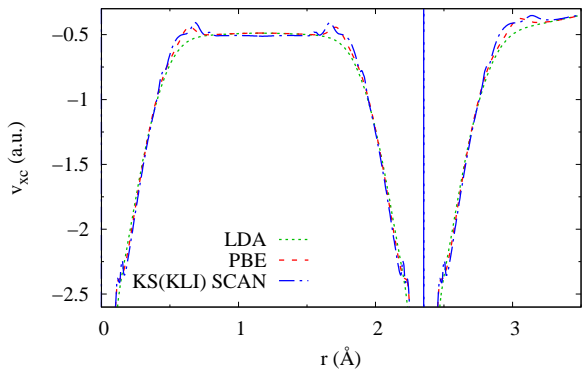}
\caption{(color online) Comparison of $v\xc$'s of bulk Si along the Si-Si bond for SCAN, LDA, and GGA. The Si atoms are located at $r=0$ and $r=2.35$\r{A}. The vertical dashed line is a numerical artifact and does not affect the band structure.}
\label{fig:SiVxc}
\end{figure}

Figs. \ref{fig:NeVxc} and \ref{fig:SiVxc} compare the OEP of SCAN meta-GGA and the LDA and the PBE xc potentials for the Ne atom and bulk Si, respectively. The exact xc potential of the Ne atom\cite{RS12} is plotted as a reference. The exact xc potential of Ne has a bump between the two shells of the charge density. The $v\xc$ of LDA does not have this feature. The $v\xc$'s of PBE and SCAN both have this bump, and they are roughly in the same position as that of the exact $v\xc$, but these $v\xc$'s are shallower than the exact $v\xc$. The $v\xc$'s of SCAN also have a few small bumps that are not in the exact $v\xc$. In the asymptotic region, the exact $v\xc$ decays as $-1/r$, and the $v\xc$'s of all the semi-local functionals decay exponentially. Fig. \ref{fig:NeVxc} shows that the $v\xc$ of SCAN has the same decay as that of LDA and PBE. For periodic systems, there is no asymptotic region, and the approximated $v\xc$'s of bulk Si in Fig. \ref{fig:SiVxc} can be expected to be closer to the exact $v\xc$ than those in Fig. \ref{fig:NeVxc}. In Fig. \ref{fig:SiVxc}, the $v\xc$ of SCAN are similar to the $v\xc$ of PBE, and they only differ in small details. Similar to Fig. \ref{fig:NeVxc}, the $v\xc$'s of SCAN for bulk Si also have small bumps. Though the meta-GGA is a higher rung functional on the Jacob's ladder than the GGA, the improvement in $v\xc$ is small going from GGA to meta-GGA, unlike going from LDA to GGA. The differences between the KS meta-GGA gaps and the PBE gaps in Table \ref{table:gap} are small as a consequence.

\begin{table*}[htbp]
\footnotesize
\begin{tabular}{lcccccccccccccc}
\hline\hline
Material & Exp. & LDA & PBE & HSE & \multicolumn{3}{c}{SCAN} & \multicolumn{3}{c}{MS2} & \multicolumn{3}{c}{MVS} & TPSS\\
 & & & & & gKS & KS(KLI) & $\Delta\xc$ & gKS & KS(KLI) & $\Delta\xc$ & gKS & KS(KLI) & $\Delta\xc$ & gKS\\
\hline
Si & 1.17 & 0.60 & 0.71 & 1.11 & 0.97 & 0.78 & 0.19 & 1.20 & 0.80 & 0.41 & 1.04 & 0.72 & 0.32 & 0.80\\
InP & 1.42 & 0.50 & 0.72 & 1.52 & 1.06 & 0.77 & 0.29 & 1.14 & 0.81 & 0.34 & 1.99 & 1.09 & 0.89 & 0.86\\
GaAs & 1.52 & 0.30 & 0.53 & 1.41 & 0.8 & 0.45 & 0.34 & 0.94 & 0.48 & 0.46 & 2.15\footnotemark[1] & 1.26 & N/A & 0.68\\
BAs & 1.60\footnotemark[4] & 1.21 & 1.26 & 1.71 & 1.51 & 1.32 & 0.19 & 1.63 & 1.33 & 0.30 & 1.56 & 1.16 & 0.40 & 1.26\\
CdSe & 1.73 & 0.44 & 0.71 & 1.66 & 1.10 & 0.76 & 0.33 & 1.06 & N/A\footnotemark[2] & N/A & 2.14 & 0.50 & 1.64 & 0.92\\
BP & 2.10 & 1.36 & 1.43 & 1.79 & 1.74 & 1.52 & 0.22 & 1.94 & 1.49 & 0.46 & 1.64 & 1.42 & 0.22 & 1.48\\
GaP & 2.35 & 1.53 & 1.69 & 2.09 & 1.94 & 1.72 & 0.21 & 1.97 & 1.73 & 0.24 & 2.23 & 1.74 & 0.49 & 1.74\\
CdS & 2.48 & 0.96 & 1.23 & 2.27 & 1.62 & 1.20 & 0.42 & 1.60 & 1.27 & 0.33 & 2.39 & 1.50 & 0.88 & 1.43\\
$\beta$-GaN & 3.17 & 1.70 & 1.69 & 2.97 & 2.03 & 1.84 & 0.20 & 1.69 & 1.70 & -0.01 & 2.50 & 1.89 & 0.62 & 1.60\\
ZnS & 3.72 & 1.87 & 2.12 & 3.32 & 2.63 & 2.16 & 0.47 & 2.52 & 2.08 & 0.45 & 3.35 & N/A\footnotemark[2] & N/A & 2.30\\
C\footnotemark[3] & 5.50 & 4.14 & 4.17 & 4.94 & 4.58 & 4.26 & 0.32 & 4.79 & 4.16 & 0.62 & 4.15 & 4.06 & 0.09 & 4.20\\
BN & 6.20 & 4.42 & 4.53 & 5.39 & 5.04 & 4.73 & 0.30 & 5.01 & 4.56 & 0.45 & 5.14 & 4.58 & 0.56 & 4.54\\
CaO & 6.93 & 3.62 & 3.75 & 5.30 & 4.29 & 3.78 & 0.50 & 4.13 & N/A\footnotemark[2] & N/A & 4.56 & 3.51 & 1.05 & 3.89\\
MgO & 7.90 & 4.70 & 4.74 & 6.46 & 5.62 & 4.80 & 0.82 & 5.20 & 5.47 & -0.27 & 6.05 & 4.58 & 1.47 & 4.86\\
NaCl & 8.97 & 4.70 & 5.08 & 6.42 & 5.86 & 5.25 & 0.61 & 5.85 & 6.51 & -0.66 & 6.61 & 4.68 & 1.93 & 5.43\\
LiF & 13.6 & 8.84 & 9.04 & 11.4 & 9.97 & 9.11 & 0.86 & 9.50\footnotemark[1] & 9.84 & N/A & 10.64 & 8.77 & 1.86 & 9.19\\
solid Ar & 14.3 & 8.44 & 8.92 & 10.33 & 9.91 & 8.89 & 1.02 & 9.95 & 9.36 & 0.58 & 10.98 & 9.19 & 1.80 & 9.56\\
\hline
MAE & & 2.08 & 1.90 & 0.88 & 1.41 & 1.84 &  & 1.28 & 1.63 &  & 1.13 & 1.89 &  & 1.76\\
MARE & & 0.46 & 0.40 & 0.13 & 0.27 & 0.38 &  & 0.26 & 0.35 &  & 0.19 & 0.36 &  & 0.36\\
\hline\hline
\end{tabular}
\footnotetext[1]{converged to the wrong ground state}
\footnotetext[2]{not converged}
\footnotetext[3]{diamond}
\footnotetext[4]{GW gap}
\caption{Calculated KS/gKS gaps and the derivative discontinuities ($\Delta\xc=E_g^\text{gKS}-E_g^\text{KS(KLI)}$). All energies are in eV. The experimental gaps are from Ref. \onlinecite{M04}. The BAs GW gap is from Ref. \onlinecite{SLC91}. The mean absolute errors (MAE) and mean absolute relative error (MARE) for the band gaps are listed. HSE\cite{HSE03,HSE05,HSE06} gaps calculated with the VASP\cite{KH93,KH94,KF96,KF96b,KJ99} code are also listed for reference. The PBE and SCAN gaps of Ge, CdO, and InN are vanishing, so they are not listed. Their experimental gaps are 0.74 eV, 0.84 eV, and 1.95 eV, respectively. Their HSE gaps are 0.65 eV, 0.94 eV, and 0.77 eV, respectively. The KS(KLI) TPSS results are not listed as most calculations fail to converge.}
\label{table:gap}
\end{table*}

\begin{figure}
\includegraphics[width=0.95\columnwidth]{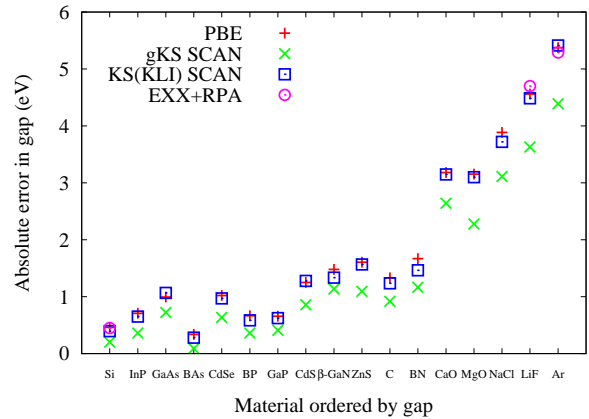}
\caption{The absolute errors of PBE, KS(KLI) SCAN, and gKS SCAN (comparing with the experimental gap) in the band gap. The EXX+RPA OEP gaps for Si, LiF and Ar, from Ref. \onlinecite{GMR06}, can be considered very close to the gaps for the corresponding exact KS potential.}
\label{fig:tabledata}
\end{figure}

\begin{figure}[htbp]
\includegraphics[width=0.95\columnwidth]{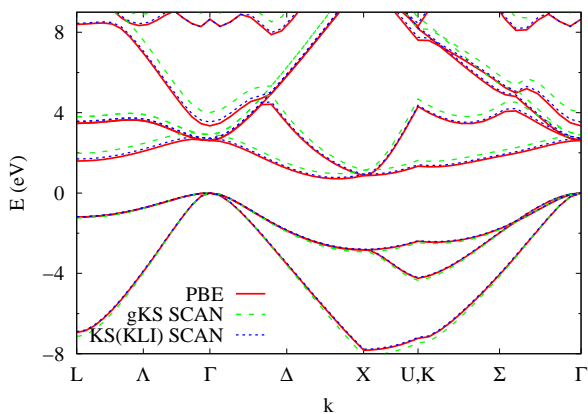}
\caption{(color online) The band structure of Si calculated with PBE, gKS SCAN, and KS(KLI) SCAN.}
\label{fig:SiBand}
\end{figure}

\begin{figure}[htbp]
\includegraphics[width=0.95\columnwidth]{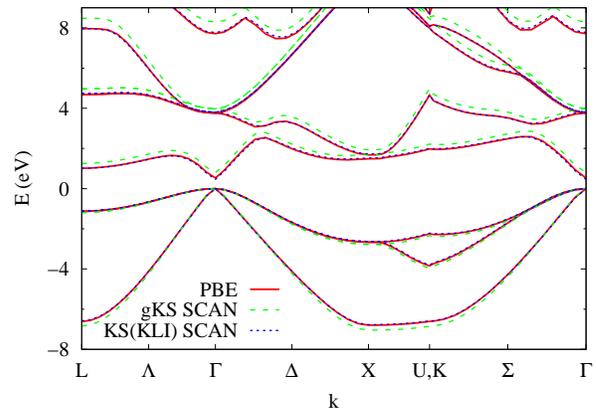}
\caption{(color online) The band structure of GaAs calculated with PBE, gKS SCAN, and KS(KLI) SCAN.}
\label{fig:GaAsBand}
\end{figure}

The band structures of Si and GaAs calculated with PBE and SCAN are plotted in Figs. \ref{fig:SiBand} and \ref{fig:GaAsBand}. The KS(KLI) SCAN band structure is very close to the PBE band structure, due to the corresponding $v\xc$ being similar to the PBE $v\xc$. The gKS SCAN band structure has the same overall shape as that of the PBE and the KS(KLI), and the main difference is in the band gap. 

Though the gKS meta-GGA band gaps improve over the PBE gaps in general, it is disappointing that gKS meta-GGA gaps for Ge, InN, and CdO still vanish. However, it is possible for meta-GGAs to open the gap for gapless materials in GGA. gKS SCAN has\cite{KPLS16} a 0.4 eV gap for $\beta\text{-MnO}_2$, which is gapless in GGA, and the value is close to the experimental value 0.3 eV. The M06L metaGGA was reported to open the gap of Ge at 0.14 eV.\cite{ZT09,PT12}

The improvement of the band gap occurs since, unlike the KS gap, the gKS gap is an approximation to the fundamental gap of the meta-GGA. A Janak-type\cite{J78} theorem has been proven for the OEP\cite{C99}, and it states that the gKS gap approximately equals the fundamental gap for the same functional, assuming fixed orbitals. This assumption does not apply to finite systems, but it is true for periodic systems, since the charge density and the orbitals of a periodic system undergo only an infinitesimal change when the number of electrons changes by one.

GGA band gaps should be compared with the OEP meta-GGA band gaps for a fair comparison between approximated functionals, since the OEP meta-GGA band gap is the KS gap. The SCAN functional is the only functional that satisfies all the known exact conditions, but the KS(KLI) SCAN gaps do not have significant improvements over the PBE gaps. This is probably due to the fact that the GGA and SCAN OEP gaps closely approximate the exact KS gap, which underestimates the fundamental gap. This has been illustrated in Fig. \ref{fig:tabledata}, where the errors of the EXX+RPA(OEP) KS gaps\cite{GMR06} are also plotted. EXX+RPA (exact exchange plus random phase approximation for correlation) is a high-level (fifth rung) method, and its OEP gaps are expected to be very close to those of the corresponding exact KS potential. Fig. \ref{fig:tabledata} shows that both PBE and KS SCAN gaps are already good approximations to the exact KS gap.

Some of the gKS band gaps of MS2 and TPSS are smaller than the corresponding OEP band gaps. We do not find this behavior in other functionals. Many of the KS(KLI) TPSS calculations fail to converge. This is probably a numerical issue in the calculation of $\nabla(\partial e\xc/\partial\tau_\sigma)$, due to the complicated functional form of TPSS.

The energy functional of the exact DFT has derivative discontinuities $\Delta\xc$ at integer electron numbers\cite{PPLB82}, where $\Delta\xc=E_g-E_g^\text{KS}$. The exact KS potential jumps up by the positive constant $\Delta\xc$ as the electron number crosses the value that makes the solid electrically neutral. LDA and GGA miss much or all of the derivative discontinuity due to the convexity of the approximated energy functional\cite{MCY08}, and they underestimate the gap as a consequence. In periodic systems, such discontinuities are only visible at band gaps. Though the derivative discontinuity has been thoroughly studied for atoms and molecules, it is difficult to obtain for periodic systems, since $I$ and $A$ cannot be calculated directly.

With the OEP meta-GGA provided in this work, we are able to estimate the $\Delta\xc$ of solids. $E_g$ can be approximated by the derivative gap $E_g^\text{deriv}=\partial E/\partial N|_{N_+}-\partial E/\partial N|_{N_-}$, which is equal to the gKS meta-GGA gap\cite{C99,MCY08,YCM12}. $E_g^\text{KS}$ is the OEP meta-GGA gap. The results are shown in Table \ref{table:gap}.

Comparison of the $\Delta\xc$'s of meta-GGAs with the exact $\Delta\xc$ is impossible for periodic systems, since no exact KS potential is available. However, Ref. \onlinecite{GSS86} provides an OEP of the GW method of bulk Si, and its $\Delta\xc$ can be seen as a good approximation to the exact $\Delta\xc$. The $\Delta\xc$ of bulk Si in Ref. \onlinecite{GSS86} is 0.58 eV, which is larger than that of all the meta-GGAs in Table \ref{table:gap}. This is not unexpected because the meta-GGA's tested in this work are not exact for $I-A$ in a solid.

\section{Real-space grid dependency}
\label{sect:grid}
\begin{table*}[htbp]
\footnotesize
\begin{tabular}{lcccccccccccccccccc}
\hline\hline
Material & solid Ar & LiF & NaCl & MgO & CaO & BN & C\footnotemark[1] & ZnS & $\beta$-GaN & CdS & GaP & BP & CdSe & BAs & GaAs & InP & Si\\
\hline
$N_r$ & 69 & 169 & 82 & 74 & 87 & 74 & 65 & 153 & 209 & 96 & 91 & 74 & 96 & 87 & 169 & 177 & 137\\
$\Delta E$ & 0.06 & 0.09 & 0.22 & 0.59 & 0.54 & 0.27 & 0.17 & 0.0008 & 0.03 & 0.17 & 0.05 & 0.03 & 0.14 & 0.01 & 0.03 & 0.06 & 0.10\\
\hline
$N_r$ & 537 & 705 & 638 & 571 & 672 & 571 & 504 & 638 & 873 & 739 & 705 & 571 & 739 & 672 & 672 & 739 & 571\\
$\Delta E$ & -0.003 & -0.01 & -0.005 & -0.002 & -0.01 & -0.02 & -0.02 & -0.006 & -0.03 & -0.0007 & -0.002 & -0.02 & -0.05 & -0.02 & -0.01 & -0.001 & -0.02\\
\hline
\hline
\end{tabular}
\footnotetext[1]{diamond}
\caption{The total energy differences between KS(KLI) SCAN and SCAN with PBE orbitals of different grids. $N_r$ is the number of radial grid points. $\Delta E=E^\text{KS(KLI)}_\text{SCAN}-E^\text{PBE orb.}_\text{SCAN}$. All energies are in eV.}
\label{table:grid}
\end{table*}

The gKS meta-GGA requires a larger real-space integration grid than GGA\cite{JBSD09}, and this is caused by the sharp variation in regions far away from nuclei of quantities containing $\tau$, such as $z=\tau^W/\tau$ used in TPSS, and $\alpha=(\tau-\tau^W)/\tau_0$ used in SCAN, MS2 and MVS, where $\tau^W=\abs{\nabla n}^2/(8n)$ is the von Wiezs\"{a}cker kinetic energy density, and $\tau_0=(3/10)(3\pi^2)^{2/3}n^{5/3}$ is the kinetic energy density of the uniform electron gas. The Becke fuzzy-cell grid is used as the real-space integration grid in most of the DFT codes. It is constructed by combining atom-centered spherical grids. The radial part of the spherical grid is dense near the nuclei, and is sparse away from the nuclei. The integration weights for the grid points in the sparse region would be larger than those in the dense region. Therefore a function with sharp features in regions away from the nuclei cannot be properly represented on the grid, and the error of integrations involving this function would be large. One needs larger grids in gKS meta-GGA calculations than those used in GGA calculations.

We find that the OEP meta-GGA is more sensitive to the real-space integration grid than the gKS meta-GGA. In the gKS case, the sensitivity to the grid shows up in the potential energy surface\cite{JWD04,JBSD09} as spurious oscillations, but the sensitivity is not obvious in a single-point calculation. In the OEP case, a single-point calculation is enough to demonstrate the grid sensitivity by comparing the total energy. The total energy of the OEP meta-GGA is variationally minimized with respect to the charge density, but the total energy of the gKS meta-GGA is variationally minimized with respect to the orbitals. Since the gKS has bigger variational degrees of freedom, the gKS total energy should always be lower than the OEP total energy, and both should be lower than the meta-GGA total energy evaluated with non-variational orbitals. However, though the built-in integration grids in the BAND code are good enough for gKS calculations, they are not sufficient for an OEP calculation: the OEP total energy calculated with these grids is always higher than the total energy evaluated with PBE orbitals. For example, the best built-in grid in BAND for the Ne atom has 120 radial points, but at least 266 radial points are required for the SCAN OEP total energy to be lower than the SCAN total energy evaluated with PBE orbitals, and 504 radial points are required for the SCAN OEP total energy to converge with respect to the grid (error$<$0.01 meV). The total energy is not sensitive to the number of angular grid points. Table \ref{table:grid} lists the total energy differences using different grids.

\begin{figure}[htbp]
\includegraphics[width=0.95\columnwidth]{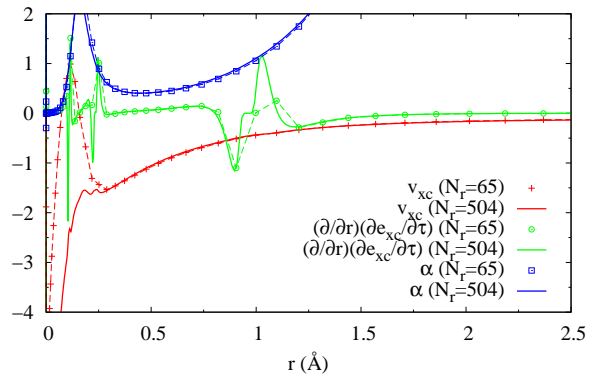}
\caption{(color online) The KS(ELP) SCAN $v\xc$ (a.u.), the radial component of $\nabla\partial e\xc/\partial\tau$ (a.u.), and $\alpha$ of the Ne atom evaluated with two different grids. The solid lines are results obtained by using 504 radial grid points, and the dots and dashed lines are results obtained by using 65 radial grid points.}
\label{fig:NeGoodBad}
\end{figure}

\begin{figure}[htbp]
\includegraphics[width=0.95\columnwidth]{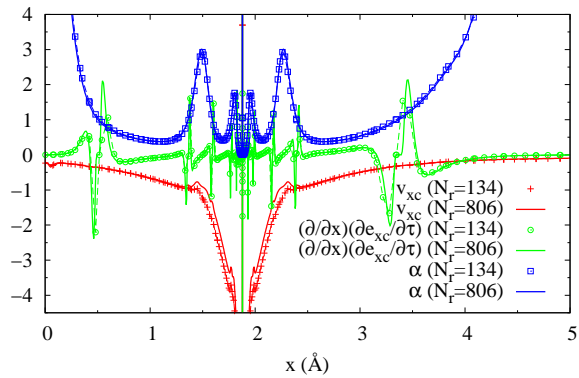}
\caption{(color online) The KS(KLI) SCAN $v\xc$ (a.u.), the $x$ component of $\nabla\partial e\xc/\partial\tau$ (a.u.), and $\alpha$ of the Ar dimer evaluated with two different grids, plotted along the Ar-Ar axis. The Ar atoms are located at $x=\pm 1.88$\r{A}, so that comparison with Ref. \onlinecite{JBSD09} is possible. Since the system is symmetric, only the $x\in[0,4]$ part is plotted. The solid lines are results obtained by using 806 radial grid points, and the dots and dashed lines are results obtained by using 134 radial grid points.}
\label{fig:Ar2GoodBad}
\end{figure}

\begin{figure}[htbp]
\includegraphics[width=0.95\columnwidth]{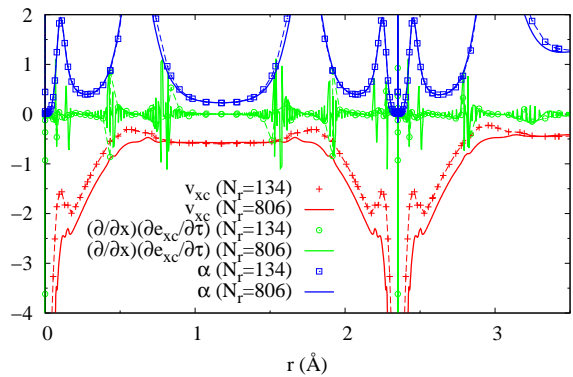}
\caption{(color online) The KS(KLI) SCAN $v\xc$ (a.u.), the $x$ component of $\nabla\partial e\xc/\partial\tau$ (a.u.), and $\alpha$ of bulk Si evaluated with two different grids, plotted along the Si-Si bond. The Si atoms are located at $r=0$ and $r=2.35$\r{A}. The solid lines are results obtained by using 571 radial grid points, and the dots and dashed lines are results obtained by using 134 radial grid points.}
\label{fig:SiGoodBad}
\end{figure}

The OEP $v\xc$ of the Ne atom, $\text{Ar}_2$ dimer, and bulk Si (along the Si-Si bond) are plotted in Figs. \ref{fig:NeGoodBad}, \ref{fig:Ar2GoodBad}, and \ref{fig:SiGoodBad}. The OEP $v\xc$ develops unnatural peaks close to nuclei when too few grid points are used. These peaks have a strong effect on the KS band gap and the total energy. For bulk Si with 134 points (dashed line in Fig. \ref{fig:SiGoodBad}), the system becomes gapless.

The grid-dependence of the OEP meta-GGA is caused by $\nabla\partial e\xc/\partial\tau_\sigma$ in Eq. \parref{eqn:working}. Figs. \ref{fig:NeGoodBad}, \ref{fig:Ar2GoodBad}, and \ref{fig:SiGoodBad} show that $\nabla\partial e\xc/\partial\tau_\sigma$ has very sharp oscillations both close to and away from the nucleus. The dots in these plots show the grid point locations when the calculation is done with a built-in grid in BAND. These grid points are sufficient for an gKS meta-GGA calculation since they are dense enough to properly describe the oscillations in $\alpha$, but they are insufficient for an OEP meta-GGA calculation since the peaks in $\nabla\partial e\xc/\partial\tau_\sigma$ are much narrower than those of $\alpha$.

\begin{figure}[htbp]
\includegraphics[width=0.95\columnwidth]{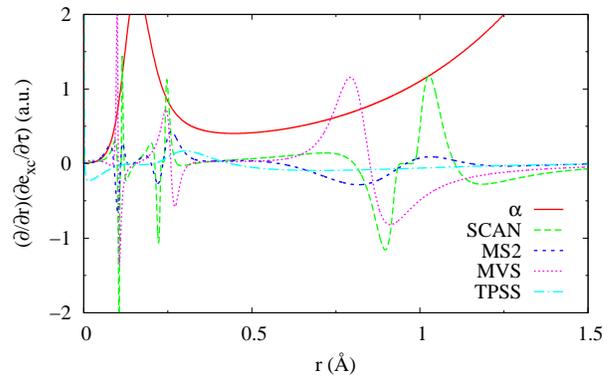}
\caption{(color online) The radial component of $\nabla\partial e\xc/\partial\tau$ of the Ne atom of different functionals. $\alpha$'s of different functionals are similar, so only that of SCAN is plotted.}
\label{fig:NeNablaDiffFunc}
\end{figure}

Fig. \ref{fig:NeNablaDiffFunc} shows $\nabla\partial e\xc/\partial\tau$ of the Ne atom for SCAN, TPSS, MS2, and MVS functionals. Though $\nabla\partial e\xc/\partial\tau$ of all functionals have oscillations, the sharpness of the peaks is different, and consequently the numbers of grid points required for convergence are also different: SCAN has very sharp peaks, and it requires 352 points to converge the total energy within 1\% error; The peaks in MVS are broader, and it only requires 252 points to converge with the same error criterion.

The oscillations in $\nabla\partial e\xc/\partial\tau$ for SCAN, MS2 and MVS are centered at or close to $\alpha=1$ in Fig. \ref{fig:NeNablaDiffFunc}. All these functionals use $\alpha$ to incorporate the kinetic energy density into the functional, and their energy densities all have the form $e\xc=e_{{\sss xc},1}+f(\alpha)(e_{{\sss xc},0}-e_{{\sss xc},1})$, where $e_{{\sss xc},0}$ and $e_{{\sss xc},1}$ are $e\xc$ constructed for $\alpha=0$ and $\alpha=1$, and $f(\alpha)$ is an interpolation function that decreases monotonically with $\alpha$ from 1 at $\alpha=0$ to 0 at $\alpha=1$ to negative values for $\alpha>1$. The oscillation in $\nabla\partial e\xc/\partial\tau$ then implies a peak in $df(\alpha)/d\alpha$. $\alpha=1$ corresponds to the uniform electron gas limit, and the $f(\alpha)$ of SCAN and MS2 are constructed to have $d^2f(\alpha)/d\alpha^2=0$ there to recover the gradient expansion approximation\cite{SXR12}. Therefore $df(\alpha)/d\alpha$ has a flat- or linearly-topped peak at $\alpha=1$, which explains the oscillation. For MVS, $d^2f(\alpha)/d\alpha^2$ vanishes at $\alpha=0.77$, and its oscillation in $\nabla\partial e\xc/\partial\tau$ occurs there. By choosing other functional forms for $f(\alpha)$, it is possible to get rid of the oscillations and the grid sensitivity. TPSS does not have this feature since it uses $z$ instead of $\alpha$.

\begin{figure}[htbp]
\includegraphics[width=0.95\columnwidth]{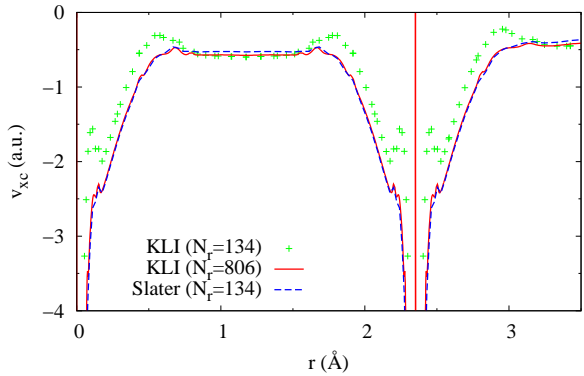}
\caption{(color online) KS(KLI) SCAN $v\xc$ of bulk Si along the Si-Si bond. The Si atoms are located at $r=0$ and $r=2.35$\r{A}.}
\label{fig:SiSlaterVxc}
\end{figure}

Even though a small grid cannot properly represent the oscillations in $\nabla\partial e\xc/\partial\tau$, Fig. \ref{fig:SiSlaterVxc} shows that the $v\xc$ of the Slater approximation (which contains $\nabla\partial e\xc/\partial\tau$ explicitly) evaluated on a small grid is already similar to the KLI $v\xc$ evaluated on a big grid. The KLI approximation is supposed to be an improvement over the Slater approximation, but it introduces big errors in $v\xc$ when using a small grid. The grid sensitivity actually enters the OEP indirectly through the $I$ integrals of Eq. \parref{eqn:Iint}. Since the Slater term contains $\nabla\partial e\xc/\partial\tau$, the $I$ integrals cannot be done accurately with a small grid.

The gKS meta-GGA potential operator in Eq. \parref{eqn:nonlocal} also contains $\nabla\partial e\xc/\partial\tau$, but the grid sensitivity of gKS is much lower than that of OEP. There is no contradiction since $\nabla\partial e\xc/\partial\tau$ does not have to be evaluated directly in gKS meta-GGA. Using integration by parts, the matrix elements $\matelem{\psi_{i\bk\sigma}}{\hat{v}\xcsigma^\text{gKS}}{\psi_{j\bk\sigma}}$ becomes
\begin{multline}
-\frac{1}{2}\int\intd^3r\psi^*_{i\bk\sigma}(\br)\nabla\cdot\left[\frac{\partial e\xc}{\partial\tau_\sigma}(\br)\nabla\psi_{j\bk\sigma}(\br)\right]\\
=\frac{1}{2}\int\intd^3r\frac{\partial e\xc}{\partial\tau_\sigma}(\br)\nabla\psi^*_{i\bk\sigma}(\br)\cdot\nabla\psi_{j\bk\sigma}(\br),
\end{multline}
and $\partial e\xc/\partial\tau$ is less sensitive to the grid than $\nabla\partial e\xc/\partial\tau$. The grid sensitivity of the OEP is also expected to show up in the time-dependent density-functional theory (TDDFT)\cite{RG84} with a meta-GGA xc kernel.

Aside from the grid problem, we find that the OEP meta-GGA in general needs more self-consistent-field (SCF) cycles to reach convergence than the gKS meta-GGA. For small gap materials, the OEPs of some meta-GGA functionals do not converge, while the corresponding gKS calculations converge normally. We tested SCAN, MS2, MVS, and TPSS functionals, and SCAN has the least convergence problem.

\section{Conclusion}
In conclusion, we implemented the ELP and the KLI approximations for the OEP meta-GGA of periodic systems, with which we study the meta-GGA band gaps of 20 semiconductors and insulators. Comparing with the GGA band gaps, the new SCAN meta-GGA in a generalized Kohn-Sham scheme is found to improve the band gaps over the GGA band gaps. The non-empirical SCAN meta-GGA outperforms the non-empirical PBE GGA, not only for diversely-bonded materials near equilibrium\cite{SRP15}, but also for the band gaps of solids. The improvement is achieved without using the expensive exact exchange, as in fourth-rung hybrid functionals. For materials wrongly predicted to be gapless in GGA, the result\cite{KPLS16} of $\beta\text{-MnO}_2$ with SCAN shows that meta-GGAs can open the gap.

Consider the ratio of the calculated to the experimental energy gap. For the 17 solids of Table \ref{table:gap}, this ratio varies from 0.35 to 0.79 with an average of 0.60 for the PBE GGA, from 0.53 to 0.94 with an average of 0.73 for the SCAN meta-GGA, and from 0.72 to 1.07 with an average of 0.89 for the HSE hybrid functional. The ratio improves uniformly from PBE to SCAN to HSE. While the hybrid functional is more accurate for the  gap than SCAN is, it is also more empirical and more computationally expensive.

For periodic systems, the OEP or ungeneralized Kohn-Sham $v\xc$'s of meta-GGAs are close to the GGA $v\xc$'s, and they only differ in small details. Consequently the OEP meta-GGA gaps (like the OEP EXX+RPA gaps where available) are not improved significantly over the GGA gaps, and the band structures of OEP meta-GGAs are similar to those of GGAs. We think it is likely that the band gap and band structure in the exact Kohn-Sham potential are close to those of GGA and OEP meta-GGA (and of the OEP hybrid) in periodic systems. Aside from the band gap, the gKS meta-GGA band structures are also similar to the GGA band structures.

Due to the sharp features in $\nabla\partial e\xc/\partial\tau_\sigma$, the OEP meta-GGA is sensitive to the real-space grid used in computation, more so than the gKS meta-GGA. Different meta-GGAs have different requirements for the minimal grid, and in general SCAN needs the biggest grid to converge of the meta-GGAs tested in this work. TDDFT with meta-GGA xc kernels is also expected to show this grid sensitivity. It is possible to avoid the grid sensitivity in the design of the functional, although this may interfere with the satisfaction of some exact conditions.

\section*{Acknowledgements}
ZY, JS and JPP are supported by the National Science Foundation(Grant No. DMR-1305135). HP, JS, and JPP are also supported by the Center for the Computational Design of Functional Layered Materials, an Energy Frontier Research Center (EFRC) funded by the U.S.\, Department of Energy, Office of Science, Basic Energy Sciences under Award No.\,DE-SC0012575. The EFRC grant contributed to the computational aspects of this work. We thank Kieron Burke for suggestions on the manuscript.

\bibliography{mggaoep}
\end{document}